%% file: p146.tex
\documentclass[11pt,epsf]{article}
\usepackage{amsmath}
\usepackage{amsfonts}
\usepackage{amssymb}
\usepackage{graphicx}
\usepackage{color}

\newcommand {\eqd} {\stackrel{\Delta} {=}}

\newcommand {\bE} {\mbox{\boldmath $E$}}

\newcommand{\calI}{{\cal I}}

\newcommand{\calU}{{\cal U}}

\newcommand{\calX}{{\cal X}}
\newcommand{\calY}{{\cal Y}}

\begin{document}
\thispagestyle{empty}
\title{Data Processing Inequalities
Based on a Certain Structured Class of Information Measures
with Application to Estimation Theory
\thanks{This research was supported by the Israeli Science Foundation (ISF),
grant no.\ 208/08.}
}
\author{Neri Merhav
}
\date{}
\maketitle

\begin{center}
Department of Electrical Engineering \\
Technion - Israel Institute of Technology \\
Haifa 32000, ISRAEL \\
{\tt merhav@ee.technion.ac.il}\\
\end{center}
\vspace{1.5\baselineskip}
\setlength{\baselineskip}{1.5\baselineskip}

\begin{abstract}
We study data processing inequalities that are derived from
a certain class of generalized information measures, 
where a series of convex functions and multiplicative likelihood ratios are nested
alternately. While these information measures can be viewed as a special case
of the most general Zakai--Ziv generalized information measure, this special nested structure
calls for attention and motivates our study. Specifically, a certain choice of the convex
functions leads to an information measure that extends the notion of the
Bhattacharyya distance (or the Chernoff divergence): While the ordinary
Bhattacharyya distance is based on the (weighted) geometric mean of two replicas of the
channel's conditional distribution, the more general information measure allows
an arbitrary number of such replicas. We apply the
data processing inequality induced by this information measure to a detailed
study of lower
bounds of parameter estimation under additive white Gaussian 
noise (AWGN) and show that in certain cases, tighter bounds
can be obtained by using more than two replicas. While the resulting lower
bound may not compete favorably with the best bounds available for the
ordinary AWGN channel,
the advantage of the new lower bound, relative to the other bounds, becomes
significant in the presence of channel uncertainty, like unknown fading.
This different behavior in the presence of channel uncertainty is explained by
the convexity property of the information measure.\\

\vspace{0.2cm}

\noindent
{\bf Index Terms:} Data processing inequality, Chernoff divergence,
Bhattacharyya distance, Gallager function, parameter estimation, fading.
\end{abstract}

\newpage
\section{Introduction}

In classical Shannon theory, data processing inequalities (in various forms)
are frequently used to prove converses to coding theorems and to establish
fundamental properties of information measures, like the entropy, the mutual
information, and the Kullback--Leibler divergence \cite{CT06}. A very well--known
example is the converse to the joint source--channel coding theorem, which
sets the stage for the separation theorem of Information Theory: When a source
with rate--distortion function $R(D)$ is encoded and transmitted across a
channel with capacity $C$, the distortion of the reconstruction at the decoder
must obey the inequality $R(D)\le C$, or equivalently, $D\ge R^{-1}(C)$.
This lower bound is achievable (e,g., by separate source coding and channel
coding) in the limit of large block length.

Ziv and Zakai \cite{ZZ73} (see also Csisz\'ar
\cite{Csiszar63}, \cite{Csiszar72}, \cite{CS04} for related work)
have observed that in order to obtain a wider class
of data processing inequalities, the (negative) logarithm function, that plays a role
in the classical mutual information, can be replaced by an arbitrary convex
function $Q$, provided that it obeys certain regularity conditions. This
generalized mutual information, $I_Q(X;Y)$, was further generalized in \cite{ZZ75} to be
based on multivariate convex functions, as opposed to the univariate convex
functions in \cite{ZZ73}. In analogy to the classical converse to the joint
source--channel coding theorem, one can then define a generalized
rate--distortion function $R_Q(D)$ (as the minimum of the generalized mutual
information between 
the source and the reproduction, s.t.\ some distortion constraint) and a generalized channel
capacity $C_Q$ (as the maximum generalized mutual information between the channel
input and output) and establish another lower bound on the distortion via the
inequality $R_Q(D)\le C_Q$ that stems from the data processing inequality of
$I_Q$. While this lower bound obviously cannot
be tighter than its classical counterpart in the limit of long blocks (which
is asymptotically achievable), Ziv and
Zakai have demonstrated that for short block codes (e.g., codes of block
length $1$), sharper lower bounds can certainly be obtained (see also
\cite{Merhav11a} for more recent developments).

Gurantz, in his M.Sc.\ work \cite{Gurantz74} (supervised by Ziv and Zakai),
continued the work in \cite{ZZ73} at a specific direction: He constructed a
special class of generalized information functionals defined by iteratively alternating
between applications of
convex functions and multiplications by likelihood ratios\footnote{The exact form
of this will be given in the sequel.} (or more generally, Radon--Nykodim
derivatives). After proving that this 
functional obeys a data processing inequality, Gurantz
demonstrated how it can be used to improve on the Arimoto bound for coding
above capacity \cite{Arimoto73} and on the Gallager upper bound 
of random coding \cite{Gallager68} by a pre-factor of $1/2$.

Motivated by the belief that the interesting nested structure 
of Gurantz' information functional can be further exploited,
we continue, in this work, to investigate this information measure and
we further study its properties and potential.

We begin by putting the Gurantz' functional in the broader perspective
of the other information measures due to Ziv and Zakai \cite{ZZ75},
\cite{ZZ73} (Section 2). Specifically,
we first discuss two possible methods to define a generalized mutual 
information from the Gurantz' functional, each one with its advantages and
disadvantages. We then show that both of these generalized mutual informations
can be viewed as special cases of
the generalized mutual information of \cite{ZZ75}, which is based on
multivariate convex functions. The proof of this fact then naturally
suggests a way to broaden the scope and define a family of information
measures with a tree structure of convex functions and likelihood ratios.

We then focus on a concrete choice of the convex
functions (Section 3) in the Gurantz' information measure (in particular,
power functions),
which turn out to yield an information measure that extends the notion of the
Bhattacharyya distance (or the Chernoff divergence): While the
ordinary Bhattacharyya distance is based on the (weighted) geometric mean of two
replicas of the channel's conditional distribution (see, e.g., \cite[eq.\
(2.3.15)]{VO79}), 
the more general information measure considered here, allows
an arbitrary number of such replicas. This generalized Bhattacharyya distance 
is also intimately related to the Gallager function $E_0(\rho,Q)$
\cite{Gallager68}, \cite{VO79}, which is indeed another information measure
obeying a data processing inequality \cite[Proposition 2]{KS93}, 
since it is yet another special case of the information measures in \cite{ZZ75}.

Finally, we apply the
data processing inequality, induced by the above described 
generalized Bhattacharyya distance, to a detailed
study of lower
bounds on parameter estimation under additive white Gaussian
noise (AWGN) and show that in certain cases, tighter bounds
can be obtained by using more than two replicas (Section 4). 
In this particular case, it turns out that three is the optimum number of
replicas in the high SNR regime.
While the resulting lower
bound may still not compete favorably with the best available bounds for the
ordinary AWGN channel,
the advantage of the new lower bound, relative to the other bounds, becomes
apparent in the presence of channel uncertainty, like in the case of an
AWGN channel with unknown fading.
This different behavior, in the presence of channel uncertainty, is explained by
the convexity property of the information measure. 

\section{Preliminaries and Basic Observations}

In \cite{Gurantz74}, a generalized information functional was defined in the
following manner:
Let $X$ and $Y$ be random variables taking on values in alphabets
$\calX$ and $\calY$, respectively, where here and throughout the sequel,
all alphabets may either be finite, countably infinite, or uncountably
infinite, like intervals or the entire real line. 
Let $x_1,x_2,\ldots,x_k$ be a
given list of symbols (possibly with repetitions) from $\calX$.
Let $Q_1,Q_2,\ldots,Q_k$ be a collection of univariate functions, defined on
the positive reals, with the following properties, holding for all $i$:
\begin{enumerate}
\item $\lim_{t\to 0} tQ_i(1/t)=0$.
\item $|Q_i(0)| < \infty$. 
\item Either the function $\hat{Q}_i\eqd Q_1\circ Q_2\circ \ldots \circ Q_i$ is monotonically
non-decreasing and $Q_{i+1}$ is convex, or $\hat{Q}_i$ is monotonically non--increasing
and $Q_{i+1}$ is concave (here, the notation $\circ$ means function
composition). 
\end{enumerate}
Now, define the {\it Gurantz' functional} as
\begin{eqnarray}
\label{gurantz1}
G(Y|x,x_1,\ldots,x_k)&=&\int_{\calY}\mbox{d}y\cdot
P_{Y|X}(y|x)\times\nonumber\\
& &Q_1\left(\frac{P_{Y|X}(y|x_1)}{P_{Y|X}(y|x)}\cdot
Q_2\left(\frac{P_{Y|X}(y|x_2)}{P_{Y|X}(y|x_1)}\cdot
Q_3\left(\ldots
Q_k\left(\frac{P_{Y|X}(y|x_k)}{P_{Y|X}(y|x_{k-1})}\right)\ldots\right)\right)\right),\nonumber
\end{eqnarray}
where here and throughout, it is understood
that integrals and probability density functions 
should be replaced, in the countable alphabet case, by summations and
probability mass functions, respectively.

The data processing inequality associated with the Gurantz' functional is the following: Let
$X\to Y\to Z$ be a Markov chain and let
$Q_1$ be a convex function which,
together with
$Q_2,\ldots,Q_k$, complies with rules 1--3 above. Then,
\begin{equation}
G(Y|x,x_1,\ldots,x_k)\ge 
G(Z|x,x_1,\ldots,x_k).
\end{equation}
The direct proof of this inequality is fairly straightforward \cite{Gurantz74}:
First, observe that 
\begin{equation}
\label{gyz}
G(Y|x,x_1,\ldots,x_k)=G(Y,Z|x,x_1,\ldots,x_k) 
\end{equation}
due to the
Markov property. Then, one can easily obtain 
a sequence of lower bounds on the right--hand--side (r.h.s.) of eq.\
(\ref{gyz}) by
successive applications of Jensen's inequality, where at each stage,
the expectation with respect to (w.r.t.) $P_{Y|X_i,Z}$ propagates into the next convex function
and then partially cancels out with the factor
$P_{Y,Z|X_i}(y,z|x_i)$ at the denominator of the likelihood ratio.

Note that according to the definition of $G(Y|x,x_1,\ldots,x_k)$, 
$x$ is the random variable that controls the distribution of $Y$ (as the
averaging is w.r.t.\ $P_{Y|X}(\cdot|x)$), whereas $x_1,\ldots,x_k$ can be viewed as
`dummy' variables. 
One way to define a generalized mutual information based on $G$, which
is a functional of $\{P_{XY}(x,y)\}$, is by
assigning a certain probability distribution to $(x,x_1,\ldots,x_k)$. 
Let $P(x,x_1,\ldots,x_k)=P_X(x)P(x_1,\ldots,x_k|x)$, where $P_X(\cdot)$ is the
actual distribution of the random variable $X$ and $P(x_1,\ldots,x_k|x)$ is an
arbitrary conditional distribution of $(X_1,\ldots,X_k)$
given $X=x$, for example, $P(x_1,\ldots,x_k|x)=\prod_{i=1}^k P_X(x_i)$ or
$P(x_1,\ldots,x_k|x)=\prod_{i=1}^k \delta(x_i-f_i(x))$ for some
deterministic functions $\{f_i\}$.
Now, for a given choice of $\{P(x_1,\ldots,x_k|x)\}$, the {\it Gurantz' mutual
information} $I_G(X;Y)$ can be defined as
\begin{equation}
I_G(X;Y)=\bE G(Y|X,X_1,\ldots,X_k)
\end{equation}
where the expectation is w.r.t.\ the above defined joint distribution of the
random variables $X$, $X_1$,..., $X_k$. This generalized mutual information
is now a well--defined functional
of $P_{XY}=P_X\times P_{Y|X}$. In principle, one may apply the generalized
data processing inequality
$I_G(X;Y)\ge I_G(X;Z)$ for any given choice of
$\{P(x_1,\ldots,x_k|x)\}$ (consider these as parameters) and then optimize the resulting
distortion bound w.r.t.\ the choice of these parameters.

Our first observation is that $I_G(X;Y)$ is a special case of
the Zakai--Ziv generalized mutual information \cite{ZZ75}, defined as
\begin{equation}
I_{ZZ}(X;Y)=\bE Q\left(\frac{\mu_1(X,Y)}{P_{XY}(X,Y)},\ldots,
\frac{\mu_k(X,Y)}{P_{XY}(X,Y)}\right),
\end{equation}
where $Q$ is a multivariate convex function of $k$ variables and
$\mu_i(\cdot,\cdot)$, $i=1,2,\ldots,k$,
are arbitrary measures on $\calX\times \calY$. 

To see why this is true,
consider the following: For each convex (resp., concave) function $Q_i(t)$, define the bivariate
{\it perspective function} $\tilde{Q}_i(s,t)=s\cdot Q_i(t/s)$, where $s > 0$, which
is a convex (resp., concave) function as well, and jointly in both variables
\cite[Subsection 3.2.6]{BV04}. Thus,
\begin{eqnarray}
& &G(Y|x_1,\ldots,x_k)\nonumber\\
&=&\int_{\calY}\mbox{d}y
P_{Y|X}(y|x)Q_1\left(\frac{P_{Y|X}(y|x_1)}{P_{Y|X}(y|x)}Q_2\left(\ldots\right)\right)\nonumber\\
&=&\int_{\calY}\mbox{d}y\cdot
P_{Y|X}(y|x')\frac{P_{Y|X}(y|x)}{P_{Y|X}(y|x')}
Q_1\left(\frac{P_{Y|X}(y|x_1)/P_{Y|X}(y|x')}{P_{Y|X}(y|x)/P_{Y|X}(y|x')}
Q_2\left(\ldots\right)\right)\nonumber\\
&=&\int_{\calY}\mbox{d}y\cdot
P_{Y|X}(y|x')\tilde{Q}_1\left(\frac{P_{Y|X}(y|x)}{P_{Y|X}(y|x')},
\frac{P_{Y|X}(y|x_1)}{P_{Y|X}(y|x')}
Q_2\left(\ldots\right)\right)\nonumber\\
&=&\int_{\calY}\mbox{d}y\cdot
P_{Y|X}(y|x')\tilde{Q}_1\left(\frac{P_{Y|X}(y|x)}{P_{Y|X}(y|x')},
\tilde{Q}_2\left(\frac{P_{Y|X}(y|x_1)}{P_{Y|X}(y|x')},\frac{P_{Y|X}(y|x_2)}{P_{Y|X}(y|x')}Q_3\left(
\ldots\right)\right)\right)\nonumber\\
&=&\ldots\nonumber\\
&=&\int_{\calY}\mbox{d}y\cdot
P_{Y|X}(y|x')\tilde{Q}_1\left(\frac{P_{Y|X}(y|x)}{P_{Y|X}(y|x')},
\tilde{Q}_2\left(\ldots
\tilde{Q}_k\left(\frac{P_{Y|X}(y|x_{k-1})}{P_{Y|X}(y|x')},
\frac{P_{Y|X}(y|x_k)}{P_{Y|X}(y|x')}\right)\ldots\right)\right)
\label{specialcase}
\end{eqnarray}
Now, under the assumed properties of the functions $\{Q_i\}$, it is easy to
see that 
\begin{equation}
\label{hatQ}
\hat{Q}(t_0,t_1,\ldots,t_k)\eqd \tilde{Q}_1(t_0,\tilde{Q}_2(t_1,\tilde{Q}_3(t_2,\ldots
\tilde{Q}_k(t_{k-1},t_k)\ldots)))
\end{equation}
is jointly convex in
$(t_0,t_1,\ldots,t_k)$. Thus, upon taking the expectation of the
last line of (\ref{specialcase}) w.r.t.\ $P_X(x')$, we have 
(after multiplying the numerator and the denominator of each likelihood ratio
by $P_X(x')$) that $\bE G(Y|X,x_1,\ldots,x_k)$ is an instance of $I_{ZZ}(X;Y)$
for every given $(x_1,\ldots,x_k)$, with the assignments
$\mu_i(x,y)=P_X(x)P_{Y|X}(y|x_i)$, $i=1,2,\ldots,k$.

\begin{figure}[ht]
\hspace*{5cm}\input{factortree.pstex_t}
\caption{\small The factor graph that represents the generalized mutual
information of eq.\ (\ref{xample}).}
\label{factree}
\end{figure}
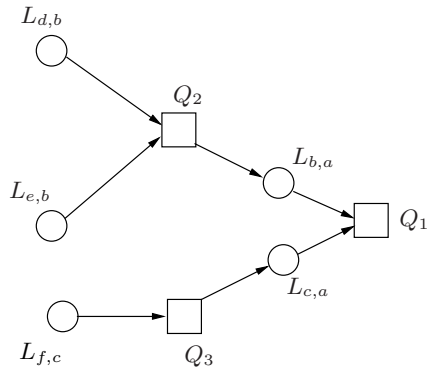

We can represent the general structure of information functionals, such as $I_G$
and $I_{ZZ}$, as well as the forms in the different lines of eq.\
(\ref{specialcase}), graphically, in
terms of {\it factor trees} (i.e., factor graphs which are trees) that obey the
following rules.
\begin{enumerate}
\item There are two types of nodes, variable nodes and function nodes, and each edge
of the tree connects a variable node and a function node.
\item The root of the tree is a function node whereas the leaves are variable nodes.
\item Each function node is
represented by a convex function $Q_i$ and each variable node is
represented by a likelihood ratio $p(y|x_k)/p(y|x_l)$, whose
shorthand notation here will be $L_{k,l}$. 
\item There is a directed edge
from function node $Q_i$ to variable node $L_{j,k}$ 
(denoted $Q_i\to L_{j,k}$)
if the information measure includes a
product of the form $Q_i(\cdot)\cdot L_{j,k}$. 
\item There is a directed
edge from variable node $L_{i,j}$ to function node $Q_k$ 
(denoted $L_{i,j}\to Q_k$)
if $L_{i,j}$ multiplies an
argument of $Q_k$. 
\item For every path $L_{i,j}\to Q_k\to L_{l,m}$, 
$j$ must be equal to $l$ (namely, $x_j=x_l$).
\item For all direct offsprings of the root, $\{L_{i,j}\}$, the second
subscript $j$ is the same.
\end{enumerate}
Now observe that $I_G$ and 
$I_{ZZ}$ correspond to two
extreme cases: While $I_{ZZ}$ corresponds to a factor tree where
all $k$ leaves are connected directly to the root,
$I_G$ corresponds to a
simple chain (i.e., every node has one offspring and there is only one leaf),
which alternates between variable nodes and function nodes. The form
that appears in the last line of (\ref{specialcase}) corresponds to a binary
tree with a comb structure, i.e., every node that is not a leaf has two
offsprings, one of which is a leaf. More generally, every factor graph with a tree
structure, that complies with the above rules, 
corresponds to a valid information measure that satisfies a data
processing inequality. For example, the factor graph of Fig.\ \ref{factree}
corresponds to the information measure
\begin{equation}
\label{xample}
\int_{\calY}\mbox{d}y\cdot p(y|x_a)
Q_1\left(\frac{p(y|x_b)}{p(y|x_a)}Q_2\left(\frac{p(y|x_d)}{p(y|x_b)},
\frac{p(y|x_e)}{p(y|x_b)}\right),\frac{p(y|x_c)}{p(y|x_a)}Q_3\left(\frac{p(y|x_f)}{p(y|x_c)}\right)\right).
\end{equation}

In view of the observation that $\bE G(Y|X,x_1,\ldots,x_k)$ a special case of 
the $I_{ZZ}(X;Y)$, there is another way to use it to obtain data processing
inequalities for communication systems. According to \cite[Theorems 3.1 and
5.1]{ZZ75}, the following is
true: Let $U\to X\to Y$ be a Markov chain and let $V=g(Y)$ where $g$ is a deterministic
function. Let $\mu_i(x,y)$, $i=1,2,\ldots,k$, be arbitrary
measures and define
$\mu_i(u,y)=P_U(u)\sum_x P_{X|U}(x|u)\mu_i(x,y)/P_X(x)$, 
$\mu_i(u,v)=\sum_{y:~g(y)=v}\mu_i(u,y)$,
$i=1,\ldots,k$. Then,
\begin{equation}
I_{ZZ}(X;Y)\ge I_{ZZ}(U;V).
\end{equation}
As described informally in the Introduction, 
the maximum of the left--hand side (l.h.s.) over $P_X$ and the minimum of the
r.h.s.\ over
$P_{V|U}$ (subject to some distortion constraint)
can be thought of as generalized channel capacity and generalized rate--distortion
function, respectively, as in \cite{ZZ75}.
Now, consider the special case where $I_{ZZ}$ is based on
a multivariate convex function $\hat{Q}$ as defined in (\ref{hatQ}), where each
bivariate convex function $\tilde{Q}_i$ is the perspective
of a certain univariate convex function, i.e., $\tilde{Q}_i(s,t)=s\cdot
Q_i(t/s)$. Then by a similar argument as above (going the other direction),
we get another information measure in the spirit of Gurantz:
\begin{equation}
I_G(X;Y)=\int_{\calX\times\calY}\mbox{d}x\mbox{d}y\cdot
P_{XY}(x,y)Q_1\left(\frac{\mu_1(x,y)}{P_{XY}(x,y)}Q_2\left(\frac{\mu_2(x,y)}{\mu_1(x,y)}
\ldots Q_k\left(\frac{\mu_k(x,y)}{\mu_{k-1}(x,y)}\right)\ldots\right)\right).
\end{equation}
Since it is a special case of $I_{ZZ}(X;Y)$, then it obviously
satisfies a strong\footnote{By ``strong data processing inequality''
we use the terminology of \cite{ZZ75}, meaning that
for a Markov chain $U\to X\to Y$ and $V=g(Y)$, we have $I_G(X;Y)\ge
I_G(U;Y)\ge I_G(U;V)$.}
data processing inequality $I_G(X;Y)\ge I_G(U;V)$.
Assuming, in addition, that the encoder is given by a deterministic function $x=f(u)$, 
we can choose 
$\mu_i(x,y)=P_X(x)P_{Y|X}(y|x_i)$, where $x_i=f(u_i)$ is a specific member in
$\calX$ and then $\mu(y|u_i)=P_{Y|X}(y|f(u_i))$.
We then obtain
\begin{eqnarray}
&&\int_{\calX\times\calY}\mbox{d}x\mbox{d}y\cdot
P_{XY}(x,y)Q_1\left(\frac{P_{Y|X}(y|f(u_1))}{P_{Y|X}(y|x)}
Q_2\left(\ldots
Q_k\left(\frac{P_{Y|X}(y|f(u_k))}{P_{Y|X}(y|f(u_{k-1}))}\right)\ldots\right)\right)\nonumber\\
&\ge&\int_{\calX\times\calY}\mbox{d}u\mbox{d}v\cdot
P_{UV}(u,v)Q_1\left(\frac{P_{V|U}(v|u_1)}{P_{V|U}(v|u)}
Q_2\left(\ldots
Q_k\left(\frac{P_{V|U}(v|u_k)}{P_{V|U}(v|u_{k-1})}\right)\ldots\right)\right).
\end{eqnarray}
Multiplying both sides by $\prod_iP_U(u_i)$
and integrating over $\{u_i\}$, we get
\begin{eqnarray}
& &\bE Q_1\left(\frac{P_{Y|X}(Y|X_1)}{P_{Y|X}(Y|X)}
Q_2\left(\ldots
Q_k\left(\frac{P_{Y|X}(Y|X_k)}{P_{Y|X}(Y|X_{k-1}))}\right)\ldots\right)\right)\nonumber\\
&\ge&\bE
Q_1\left(\frac{P_{V|U}(V|U_1)}{P_{V|U}(V|U)}
Q_2\left(\ldots
Q_k\left(\frac{P_{V|U}(V|U_k)}{P_{V|U}(V|U_{k-1})}\right)\ldots\right)\right).
\end{eqnarray}
where the expectation on the l.h.s.\ is w.r.t.\
$P_{XY}(x,y)\prod_iP_X(x_i)$, 
and the expectation on the r.h.s.\ is w.r.t.\ $P_{UV}(u,v)\prod_iP_U(u_i)$.
This is different from the data processing theorem in \cite{Gurantz74}, because
it allows `moving' in both directions of the Markov chain and not only to the
right.

To summarize, we have seen two approaches to derive data processing inequalities 
from the inequality $G(Y|u,u_1,\ldots,u_k)\ge
G(V|u,u_1,\ldots,u_k)$ for a Markov chain $U\to Y\to V$ (where have slightly
changed the notation relative to eq.\ (\ref{gurantz1})):
According to the first approach, one allows an arbitrary distribution $P_{U_1,\ldots U_k|U}$ and
averages both sides w.r.t.\ $P_U\times P_{U_1,\ldots U_k|U}$. This
defines the $I_G(U;Y)$ and $I_G(U;V)$ as functionals of $P_{UY}$ and $P_{UV}$,
respectively, where 
$P_{U_1,\ldots U_k|U}$ serve as free parameters
that can be optimized, to get the tightest distortion bound.
The advantage of this approach 
is the free choice of $P_{U_1,\ldots U_k|U}$, which gives many degrees of freedom.
The disadvantage is that $I_G(U;Y)$ depends on the
source and the encoder and there is no apparent way to prove a strong data
processing theorem, in general, i.e., to prove that $I_G(U;Y)$
can be further upper bounded by $I_G(X;Y)$ 
(whatever its definition may be) 
and thereby define a channel
capacity, that is independent of the source (in addition to a generalized rate
distortion function, which is $\min I_G(U;V)$ s.t.\ some distortion
constraint). The inequality $I_G(U;Y)\ge I_G(U;V)$ is relevant to
situations where there is no encoder to be optimized, namely, when the
channel from $U$ to $Y$ is given and cannot be shaped by encoding. This
happens, for example, in parameter estimation problems.

According to the second approach, one limits $P_{U_1,\ldots U_k|U}(u_1,\ldots,u_k|u)$ to be
$\prod_{i=1}^kP_U(u_i)$. This leaves no degrees of freedom, but it admits
a strong data processing theorem, and hence allows
to define both a generalized rate--distortion function and a generalized
channel capacity,
whose calculations are completely decoupled of each other.
It is also much simpler to use.
This type of data processing inequality is more suitable for coded
communication systems, where there is also an encoder to optimize.

From this point onward, we essentially confine ourselves to the second option, mainly for
reasons of simplicity. 

\section{Choice of the Convex Functions}

An interesting and convenient choice of the functions $\{Q_i\}$ is the
following:
$Q_1(t)=-t^{a_1}$, and $Q_i(t)=t^{a_i}$ for $i \ge 2$, where $0\le a_i\le 1$, $i=1,\ldots,k$.
In this case, $\tilde{Q}_i(t)=-t^{\prod_{j=1}^ia_j}$ is monotonically
decreasing and $Q_{i+1}$ is concave, so this choice complies with the
rules. In this case, we have:
\begin{eqnarray}
\label{specG}
G(Y|x_0,x_1,\ldots,x_k)&=&-\int_{\calY}\mbox{d}y
P_{Y|X}(y|x_0)\times\nonumber\\
& &\left(\frac{P_{Y|X}(y|x_1)}{P_{Y|X}(y|x_0)}\left(
\frac{P_{Y|X}(y|x_2)}{P_{Y|X}(y|x_1)}
\left(\ldots\left(
\frac{P_{Y|X}(y|x_k)}{P_{Y|X}(y|x_{k-1})}\right)^{a_k}\right)^{a_{k-1}}\
\ldots\right)^{a_2}\right)^{a_1}\nonumber\\
&=&-\int_{\calY}\mbox{d}y \prod_{i=0}^k P_{Y|X}^{b_i}(y|x_i)
\end{eqnarray}
where $\{b_i\}$ are given by:
\begin{eqnarray}
b_0&=&1-a_1\nonumber\\
b_1&=&(1-a_2)a_1\nonumber\\
b_2&=&(1-a_3)a_1a_2\nonumber\\
&\ldots&\nonumber\\
b_{k-1}&=&(1-a_k)\prod_{i=1}^{k-1}a_i\nonumber\\
b_k&=&\prod_{i=1}^k a_i
\end{eqnarray}
Note that the coefficients $b_0,\ldots,b_k$ are all non--negative and
their sum is equal to $1$. 
Conversely, for every set of coefficients $\{b_i\}$ with these properties, one can
find $a_1,\ldots,a_k$, all in $[0,1]$, using the following inverse
transformation:
\begin{eqnarray}
a_1&=&1-b_0\nonumber\\
a_2&=&1-\frac{b_1}{1-b_0}\nonumber\\
&\ldots&\nonumber\\
a_k&=&1-\frac{b_{k-1}}{1-\sum_{i=0}^{k-2}b_i}.
\end{eqnarray}
This allows us parametrize the information measure directly in terms of
an arbitrary set of non--negative numbers $\{b_i\}$ summing to unity, without
worrying about $\{a_i\}$. The resulting information measure can then be viewed
as an extension of the Chernoff divergence between two conditional
densities, $P_{Y|X}(y|x_0)$ and $P_{Y|X}(y|x_1)$, to a general number of
densities, where the powers of $\{P_{Y|X}(y|x_i)\}$ always sum up to unity.
Specializing this to the case $b_i=1/(k+1)$ for all $i=0,1,\ldots,k$, eq.\
(\ref{specG}) extends the
Bhattacharyya distance. Following the discussion of the second option at the end of Section 2,
if, in addition, we assign 
$P_{X_1,\ldots,X_k|X_0}(x_1,\ldots,x_k|x_0)=\prod_{i=1}^kP_{X}(x_i)$,
then $I_G(X;Y)=\bE G(Y,X,X_1,\ldots,X_k)=-e^{-E_0(k,P_X)}$, where $E_0$ is the
Gallager function \cite{Gallager68}
\begin{equation}
E_0(\rho,P_X)=-\ln\left\{\int_{\calY}
\mbox{d}y\left[\int_{\calX}\mbox{d}x
P_X(x)P_{Y|X}^{1/(1+\rho)}(y|x)\right]^{1+\rho}\right\}.
\end{equation}
Thus, $I_G(X;Y)$ extends, not only the Chernoff
divergence, but also the Gallager function, albeit only at integer values of
the parameter $\rho$. Indeed, it was shown in \cite[Proposition 2]{KS93} that the
Gallager function (for every real $\rho \ge 0$) satisfies a data processing
inequality, because it
is also a special case of $I_{ZZ}(X;Y)$.
In other words, the generalized Chernoff divergence can be obtained as a
special case of $I_{ZZ}(X;Y)$ in two different
ways: one is via $I_G$ and the other is via the
Gallager function. 
The advantage of working with Gallager's function for 
integer values of $\rho$, is that an integral raised to an integer power
$(k+1)$
can be expressed in terms of $(k+1)$--dimensional integration over
the $(k+1)$ replicas, $x_0$,$x_1$,...,$x_k$, that in turn
can be commuted with the additional out--most integration over $\calY$. In some situations,
this enables explicit calculations more conveniently.

\section{Application to Estimation Theory}

In this section, we apply the data processing inequality associated with
the generalized Bhattacharyya distance to obtain a Bayesian lower bound
on the estimation error of parameter estimators of a parameter $u$
modulated in a signal $x(t,u)$ that is in turn corrupted by Gaussian white noise. 
As mentioned earlier, we essentially adopt the second approach discussed at
the end of Section 2: Although we use the data processing inequality
$I_G(U;V)\le I_G(U;Y)$, in some of our derivations, we eventually further upper bound $I_G(U;Y)$ by a
universal bound, that is independent of the modulation scheme $x(t,\cdot)$, so in
a way, it conveys the notion of generalized capacity.
The model we focus on is the following.

The source symbol $U$, which is uniformly distributed in $\calU=[-1/2,+1/2]$,
plays the role of a random parameter to be estimated.
For reasons of convenience, we define the distortion measure between a realization $u$ 
of the source
and an estimate $v$ (both in $\calU$) as
\begin{equation}
d(u,v)=[(u-v)~\mbox{mod}~1]^2.
\end{equation}
where
\begin{equation}
t~\mbox{mod}~1\eqd \left< t+\frac{1}{2}\right>-\frac{1}{2}
\end{equation}
$\left<r\right>$ being the fractional part of $r$,
that is, $\left<r\right>=r-\lfloor r\rfloor$.
Note that in the high--resolution limit (corresponding to
the high signal--to--noise (SNR) limit), the modulo 1 operation has a negligible effect, and hence
$d(u,v)$ becomes essentially equivalent to the ordinary
quadratic distortion. Indeed, most of our results in the sequel, refer to the
high SNR regime. At any rate, under the modulo 1 quadratic distortion measure,
it is convenient to visualize $U$ as
being evenly
distributed across the circumference of a circle of radius $1/(2\pi)$ (or as a 
phase parameter)
and then $d(u,v)$ is the squared length of the shorter arc (or the smaller
angel) between the two
corresponding points on the circle.

The channel is assumed to be an AWGN channel, namely, the channel output is
given by
\begin{equation}
y(t)=x(t,u)+n(t),~~~~0\le t < T,
\end{equation}
where $x(t,u)$ is an arbitrary waveform of unlimited bandwidth, parametrized by $u$
and $n(t)$ is AWGN with two--sided spectral density $N_0/2$. The energy 
\begin{equation}
E=\int_O^Tx^2(t,u)\mbox{d}t
\end{equation}
is assumed to be independent of $u$ (for reasons of simplicity).
The estimator $v$ is assumed to be a functional of the channel
output waveform $\{y(t),~0\le t < T\}$.

Before deriving lower bounds on the estimation error, $\bE d(U,V)$, we first need to derive
the generalized rate--distortion function and the generalized channel capacity
pertaining to the generalized Bhattacharyya distance.
This will be done in the next two subsections.

\subsection{Derivation of $R(D)$}

The ``rate--distortion function'' $R(D)$ w.r.t.\ the information measure under
discussion is given by the minimum of
$$I(U;V)=-\int_{-1/2}^{+1/2}\mbox{d}v\left[\int_{-1/2}^{+1/2}\mbox{d}u
P_{V|U}^{1/(k+1)}(v|u)\right]^{k+1}$$
subject to the constraints $\bE d(U,V)\le D$ and $\int_{-1/2}^{+1/2}\mbox{d}v
P_{V|U}(v|u)=1$. As explained in \cite{ZZ73}, it is enough to consider channels
of the form $P_{V|U}(v|u)=f(v-u)$. Defining $w=(v-u)~\mbox{mod}~1$,
the problem is then
equivalent to 
\begin{eqnarray}
& &\max \int_{-1/2}^{+1/2}\mbox{d}w\cdot f^{1/(k+1)}(w)\nonumber\\
& &\mbox{s.t.}~~~
\int_{-1/2}^{+1/2}\mbox{d}w \cdot w^2f(w)\le D\nonumber\\
& &~~~~~~~~\int_{-1/2}^{+1/2}\mbox{d}w\cdot f(w)=1.
\end{eqnarray}
This problem is easily solved using calculus of variations \cite{Andelman74}. Suppose that
$f^*$ is the optimum density and let $f=f^*+\delta g$, where
$g$ satisfies 
\begin{equation}
\int_{-1/2}^{+1/2}\mbox{d}w\cdot g(w)=0.
\end{equation}
Defining the Lagrangian
\begin{equation}
J(f)=-\int_{-1/2}^{+1/2}\mbox{d}w\cdot
f^{1/(k+1)}(w)+\lambda\int_{-1/2}^{+1/2}\mbox{d}w \cdot w^2f(w)
+\nu \int_{-1/2}^{+1/2}\mbox{d}w\cdot f(w),
\end{equation}
the condition for $f^*$ being an extremum is $\partial J(f+\delta
g)/\partial\delta|_{\delta=0}=0$ for all $g$.
Now,
\begin{equation}
\frac{\partial J(f+\delta g)}{\partial\delta}\bigg|_{\delta=0}=
\int_{-1/2}^{+1/2}\mbox{d}w \cdot g(w)\left[
-\frac{1}{(k+1)f^{k/(k+1)}(w)}+\lambda w^2 +\nu\right]=0.
\end{equation}
For this integral to vanish for every $g$, one must have
\begin{equation}
-\frac{1}{(k+1)f^{k/(k+1)}(w)}+\lambda w^2 +\nu=\mbox{const}.
\end{equation}
This means that $f^*$ is of the form
\begin{equation}
f^*(w)=\frac{C(s)}{(1+s w^2)^{1+1/k}},
\end{equation}
where
\begin{equation}
C(s)=\left[\int_{-1/2}^{+1/2}\frac{\mbox{d}w}{(1+s
w^2)^{1+1/k}}\right]^{-1},
\end{equation}
and the parameter $s$ is determined such that
\begin{equation}
C(s)\int_{-1/2}^{+1/2}\frac{w^2\mbox{d}w}{(1+s
w^2)^{1+1/k}}=D.
\end{equation}
Define also
\begin{equation}
F(s)=\int_{-1/2}^{+1/2}\frac{w^2\mbox{d}w}{(1+s
w^2)^{1+1/k}}.
\end{equation}
Let us denote then $D_s=C(s)F(s)$.
Then,
\begin{eqnarray}
-R(D_s)&=&\left[\int_{-1/2}^{+1/2}\mbox{d}w
[f^*(w)]^{1/(k+1)}\right]^{k+1}\nonumber\\
&=&C(s)\left[\int_{-1/2}^{+1/2}\frac{\mbox{d}w}{(1+sw^2)^{1/k}}\right]^{k+1}\nonumber\\
&=&C(s)[G(s)]^{k+1},
\end{eqnarray}
where we have defined
\begin{equation}
G(s)=\int_{-1/2}^{+1/2}\frac{\mbox{d}w}{(1+sw^2)^{1/k}}.
\end{equation}
To summarize, we have obtained a parametric representation of $R(D)$ via the
variable $s$:
\begin{eqnarray}
\label{rd}
D_s&=&C(s)F(s)\\
R(D_s)&=&-C(s)[G(s)]^{k+1},
\end{eqnarray}
For later use, we point out that the 
functions $C(s)$, $F(s)$, and $G(s)$ are intimately related. First, observe
that
\begin{eqnarray}
G(s)&=&\int_{-1/2}^{+1/2}\frac{(1+sw^2)\mbox{d}w}{(1+sw^2)^{1+1/k}}\nonumber\\
&=&\frac{1}{C(s)}+sF(s).
\end{eqnarray}
Also, using integration by parts,
\begin{eqnarray}
G(s)&=&w(1+sw^2)^{-1/k}\bigg|_{-1/2}^{+1/2}+\frac{2s}{k}\cdot F(s)\nonumber\\
&=&\left(1+\frac{s}{4}\right)^{-1/k}+\frac{2s}{k}\cdot F(s).
\end{eqnarray}
Thus,
\begin{equation}
\frac{1}{C(s)}+sF(s)=\left(1+\frac{s}{4}\right)^{-1/k}+\frac{2s}{k}\cdot F(s),
\end{equation}
which gives a direct relationship between $C(s)$ and $F(s)$ whenever $k\ne 2$.
For $k=2$, the terms pertaining to $F(s)$ cancel out, but we then have an explicit
formula for $C(s)$.

While in general, $R(D)$
is given only a parametric form and not directly,
in the limits of very low and very high distortion, one
can approximate $R(D)$ directly as an explicit function of $D$. 
In particular, it is shown in Appendix A that in the low resolution regime,
\begin{equation}
D(R)\approx\frac{1}{12}-\frac{1}{15}\sqrt{1+R},
\end{equation}
where it should be kept in mind that for this information measure, $R$ takes
on values in the interval $[-1,0]$.
Here and throughout the sequel, the notation $A\approx B$ means that
$A/B$ tends to unity as a certain parameter (in this case, $R$) tends to
a certain limit (in this case, $-1$), which will always be clear from the
context. Here, the term $1/12$ is the variance of $U$, which is uniform over
$[-1/2,+1/2]$, as no useful information is available except the prior.

In the high--resolution regime ($R\to 0$), the behavior depends
on whether $k=1$, $k=2$, or $k > 2$. In Appendix B, derivations are provided
for all three cases. For $k=1$, the rate--distortion function is approximated
as
\begin{equation}
\label{rdk=1}
R(D)\approx -4c_1\sqrt{D}.
\end{equation}
or equivalently, the distortion--rate function is
\begin{equation}
\label{drk=1}
D(R)\approx\frac{R^2}{16c_1^2}, 
\end{equation}
where
\begin{equation}
c_1=\int_{-\infty}^{+\infty}\frac{\mbox{d}t}
{(1+t^2)^2}.
\end{equation}
For $k > 2$, we have
\begin{equation}
\label{drk>2}
R(D)\approx-4\left(\frac{k}{k-2}\right)^k\cdot D~~\mbox{or}~~
D(R)\approx-\frac{1}{4}\left(1-\frac{2}{k}\right)^k\cdot R,
\end{equation}
The case $k=2$ lacks an explicit closed--form direct relation between $R$ and
$D$, but it shows that
\begin{equation}
\label{drk=2}
\log D\approx\log [-R(D)],
\end{equation}
which means that the relation between $R$ and $D$ is essentially
linear, like in the case $k >2$, but in a slightly weaker sense.
It is also easy to extend all the derivations to higher--order moments
modulo 1 (see Appendix C for the high resolution analysis).

\subsection{Derivation of $I_G(U;Y)$}

As mentioned earlier, the channel is assumed to be an AWGN channel
with unlimited bandwidth. The probability law of the channel from $U$ to $Y$
is given by
\begin{equation}
P_{Y|U}(y|u)\propto\exp\left\{-\frac{1}{N_0}\int_0^T[y(t)-x(t,u)]^2dt\right\},
\end{equation}
where $y$ in the l.h.s.\ designates the entire channel output waveform $\{y(t),~0\le t <
T\}$, and
$\propto$ means that the constant of proportionality does not depend on
$u$. Let us denote
\begin{equation}
\rho(u,u')=\frac{1}{E}\cdot\int_0^Tx(t,u)x(t,u')\mbox{d}t.
\end{equation}
Consider the integral
\begin{eqnarray}
\label{ksim}
& &\int \mbox{d}y \prod_{i=0}^k[P_{Y|U}(y|u_i)]^{1/(k+1)}\nonumber\\
&=&\bE\left\{\frac{\prod_{i=1}^k[P_{Y|U}(y|u_i)]^{1/(k+1)}}{P_{Y|U}(y|u_0)^{k/(k+1)}}\bigg|
U=u_0\right\}
\nonumber\\
&=&\bE\left\{\exp\left[\frac{k}{(k+1)N_0}\int_0^T[y(t)-x(t,u_0)]^2dt-
\frac{1}{(k+1)N_0}\sum_{i=1}^k\int_0^T[y(t)-x(t,u_k)]^2dt
\right]\bigg|U=u_0\right\}\nonumber\\
&=&\bE\exp\left\{\frac{2}{(k+1)N_0}\int_0^T[x(t,u_0)+n(t)]\left[\sum_{i=1}^kx(t,u_i)-
kx(t,u_0)\right]dt\right\}\nonumber\\
&=&\exp\left\{-\frac{E}{N_0}\left[1-
\frac{1}{(k+1)^2}\sum_{i=0}^k\sum_{j=0}^k\rho(u_i,u_j)\right]\right\},
\label{1stexp}
\end{eqnarray}
where the last passage is associated with the calculation of the
moment--generating
function of the Gaussian random variable
\begin{equation}
Z=\int_0^Tn(t)\left[\sum_{i=1}^kx(t,u_i)-kx(t,u_0)\right]\mbox{d}t
\end{equation}
which has zero mean and variance
$\frac{N_0}{2}\int_0^T[\sum_{i=1}^kx(t,u_i)-kx(t,u_0)]^2\mbox{d}t$.

The next step, in principle, is take another expectation
over the last line of (\ref{ksim}) w.r.t.\ the randomness of $\{U_i\}$.
This can be done explicitly for some specific classes of signals (e.g., when $U$ is a
phase parameter of a sinusoid), but in general, it is not a trivial task. As
in \cite{Andelman74} and \cite{ZZ75}, we then resort to a lower bound (hence an upper bound
on $I_G(U;Y)$) based on Jensen's
inequality, by raising the expectation operator to the exponent.
Denoting
\begin{equation}
\bar{x}(t)=\bE\{x(t,U)\}=\int_{-1/2}^{+1/2}\mbox{d}u\cdot x(t,u),
\end{equation}
it is easily observed that since $\{U_i\}$ 
are independent, then for all $i\ne j$:
\begin{equation}
\bE\rho(U_i,U_j)=\frac{1}{E}\cdot \bE\left\{\int_0^Tx(t,U_i)x(t,U_j)\mbox{d}t\right\}=
\frac{1}{E}\int_0^T[\bar{x}(t)]^2\mbox{d}t\eqd \varrho.
\end{equation}
Note that the parameter $\varrho$ is always between $0$ and $1$ and
it depends only on the parametric family of
signals.\footnote{For example, if $x(t,u)=x_0(t-u)$
is a rectangular pulse of duration $\Delta$ then $\varrho=\Delta/T$.}
Specifically, continuing from the last line of (\ref{1stexp}), we have
\begin{eqnarray}
& &\bE\exp\left\{-\frac{E}{N_0}\left[1-
\frac{1}{(k+1)^2}\sum_{i=0}^k\sum_{j=0}^k\rho(U_i,U_j)\right]\right\}\nonumber\\
&=&
\exp\left\{-\frac{E}{N_0}\left[1-\frac{1}{k+1}\right]\right\}\cdot\bE\exp
\left\{\frac{E}{N_0(k+1)^2}\sum_{i\ne j}\rho(U_i,U_j)\right\}\nonumber\\
&\ge&\exp\left\{-\frac{E}{N_0}\cdot\frac{k}{k+1}\right\}\cdot\exp
\left\{\frac{E}{N_0(k+1)^2}\sum_{i\ne j}\bE\rho(U_i,U_j)\right\}\nonumber\\
&=&\exp\left\{-\frac{E}{N_0}\cdot\frac{k}{(k+1)}(1-\varrho)\right\}.
\label{capawgn}
\end{eqnarray}
Note that the expression $E(1-\varrho)$, that appears in the exponent,
is equal to $\int_0^T\mbox{Var}\{x(t,U)\}\mbox{d}t$, which is a measure
of the variability, or the sensitivity of the $x(t,u)$ to the parameter $u$
(in analogy the Cram\'er--Rao bound that depends on the energy of the
derivative of the signal w.r.t.\ $u$, as 
another measure of sensitivity). Accordingly, classes of
signals with smaller values of $\varrho$ (or equivalently, higher values of
the integrated variance of $x(t,U)$) are expected to yield higher value of
$I_G(U;Y)$, and hence smaller estimation error, at least as far as our bounds predict,
and since $\varrho$ cannot be negative, the best 
classes of signals, in this sense, are those for which $\varrho=0$.
Note also that for Jensen's inequality to be reasonably tight,
the random variables $\{\rho(U_i,U_j)\}$ should be
all close to their expectation $\varrho$ with very high probability, and
if this expectation vanishes, as suggested, then $\{\rho(U_i,U_j)\}$
should all be nearly zero with very high probability. We will get back to
classes of signals with this desirable rapidly vanishing correlation property later on.

\subsection{Estimation Error Bounds for the AWGN Channel}

We now equate $R(D)$ to $I_G(U;Y)$ in order to obtain estimation error bounds
in the high SNR regime, where the high--resolution expressions of $R(D)$ are
relevant. As discussed above, in this regime, we will neglect the effect of the
modulo 1 operation in the definition of the distortion measure, and will refer
to it hereafter as the ordinary quadratic distortion measure.
The choice $k=1$ yields $I_G(U;Y)\le -e^{-(1-\varrho)E/(2N_0)}$ (see also
\cite{ZZ75}), and following eq.\ (\ref{drk=1}), this yields
\begin{equation}
\bE(U-V)^2\ge
D\left(-e^{-(1-\varrho)E/(2N_0)}\right)=\frac{e^{-(1-\varrho)E/N_0}}{16c_1^2},
\end{equation}
and so, the exponential decay of the lower bound is according to
$e^{-(1-\varrho)E/N_0}$.
For $k=2$, according to eq.\ (\ref{drk=2}), we have
$\log D \approx 2(1-\varrho)E/(3N_0)$, which means
an exponential decay according to $e^{-2(1-\varrho)E/(3N_0)}$, which is better.
For $k\ge 3$, we use (\ref{drk>2}) and the resulting bound decays
according to $\exp\{-(1-\rho)kE/[(k+1)N_0]\}$, which is better than
the result of $k=1$, but not as good as the one of $k=2$. Thus, the best
choice of $k$ for the high SNR regime is $k=2$, namely, a generalized
Bhattacharyya distance with $k+1=3$ replicas, rather the two 
replicas of the ordinary Bhattacharyya distance.

Note that since $\varrho\ge 0$, as mentioned earlier,
then for any family of signals, the exponential function
$e^{-2E/(3N_0)}$ is a {\it universal} lower bound 
(at high SNR) in the sense 
that it applies, not only
to every estimator of $U$, but also to every
parametric family of signals $\{x(t,u)\}$, i.e., to every modulation scheme
without being dependent on this modulation scheme (see also \cite{ZZ75}).
This is in contrast to most of the estimation error bounds in the literature.
In other words, it sets a fundamental limit on the entire communication
system and not only on the receiver end for a given transmitter.
Indeed, for some classes of signals, an
MSE with exponential decay in $E/N_0$ is attainable at least in the
high SNR regime, although there might be gaps in the actual exponential rates
compared to the above mentioned bound. 
For example, in \cite{Merhav11b}, it is discussed that 
in the case of time delay estimation ($x(t,u)=x_0(t-u)$),
it is possible to achieve an MSE of the exponential order 
of $e^{-E/(3N_0)}$ by allowing
the pulse $s_0(t)$ to have bandwidth that grows exponentially with
$T$.\footnote{
Other examples include chirp--like signals, e.g., $x(t,u)=\sin(u e^{Rt})$
(for some given $R > 0$),
as well as chaotic signals parametrized by their initial condition --
see \cite{Hen02}, \cite{HM04} and references therein.}
Thus, by improving the lower bound $\exp(-E/N_0)$ (a special case of the
above with $k=1$) to $\exp[-2E/(3N_0)]$, we are halving the gap between the
exponential rates of the upper bound and the lower bound, from $2E/(3N_0)$ to 
$E/(3N_0)$.

Our asymptotic lower bound should be compared to other lower bounds
available in the literature. One natural candidate would be the
Weiss--Weinstein bound (WWB) \cite{Weiss85}, \cite{WW85a}, 
\cite{WW85b}, which for the model under discussion at high
SNR, reads \cite[p.\ 66]{Weiss85}:
\begin{equation}
\mbox{WWB}=\sup_{h\ne
0}\frac{h^2\exp\{-[1-r(h)]E/(2N_0)\}}{2(1-\exp\{-[1-r(2h)]E/(2N_0)\})},
\end{equation}
where $r(h)=\rho(u,u+h)=\int_0^Tx(t,u)x(t,u+h)\mbox{d}t/E$ is assumed to depend only on
$h$ and not on $u$. While this is an excellent bound for a given modulation
scheme $\{x(t,u),~u\in\calU\}$, it does not seem to lend itself easily to the
derivation of
universal lower bounds, as discussed above. To this end, in principle, the WWB should be
minimized over all feasible correlation functions $r(\cdot)$, which is not
a trivial task. A reasonable compromise is to first minimize the WWB over
$r(\cdot)$ for a given $h$, and then to maximize the resulting expression
over $h$ (i.e., max--min instead of min--max). Since the expression of the
bound is a monotonically increasing function of both $r(h)$ and $r(2h)$,
and since both $r(h)$ and $r(2h)$ cannot 
be smaller than $-1$, we end up with
\begin{equation}
\mbox{WWB}=
\frac{e^{-E/N_0}}{2(1-e^{-E/N_0})}
\end{equation}
as a modulation--independent bound. This is a faster exponential decay rate 
(and hence weaker asymptotically) than that of our
proposed bound for $k=2$. 

It is possible, however, to obtain a 
universal lower bound stronger than both bounds by a simple
channel--coding argument, which is in the spirit of the Ziv--Zakai bound
\cite{ZZ69}. This bound is given by
(see Appendix D for the derivation):
\begin{equation}
\bE(U-V)^2\ge
\frac{1}{8M^2}\cdot Q\left(\sqrt{\frac{E}{N_0}\cdot\frac{M}{M-2}}\right),
\end{equation}
where
\begin{equation}
Q(x)\eqd \frac{1}{\sqrt{2\pi}}\int_x^\infty e^{-t^2/2}\mbox{d}t
\end{equation}
and where $M$ is a free parameter, an even integer
not smaller than $4$, which is 
subjected to optimization.
Throughout the sequel, we refer to this bound as the {\it channel--coding
bound}. In the high SNR regime, the exponential order of the channel--coding
bound (for fixed $M$) is
\begin{equation}
\exp\left\{-\frac{E}{2N_0}\cdot \frac{M}{M-2}\right\},
\end{equation}
which for large enough $M$ becomes arbitrarily close to
$e^{-E/(2N_0)}$, and hence better than the data--processing bound of
$e^{-2E/(3N_0)}$. Note that the Ziv--Zakai bound \cite{ZZ69} would be weaker in this
context of universal lower bounds, since it is based on binary hypothesis
testing ($M/2=2$), yielding an exponent of $e^{-E/N_0}$.

In view of this comparison, 
it is natural to ask then what is benefit of our data processing lower bound.
The answer is that the potential of the data--processing bound is much better exploited in
situations of channel uncertainty, like in channels with fading.
This is the subject
of the next subsection.

\subsection{Estimation Error Bounds for the AWGN Channel with Fading}

It turns out that the feature that makes the data--processing--theorem approach 
to error lower bounds more powerful,
relatively to other approaches, 
is the convexity property of the generalized mutual information (in this case,
$I_G(U;Y)$) w.r.t.\ the channel $P_{Y|U}$. 
Suppose that the channel actually depends on an additional random parameter
$A$ (independent of $U$), that is known to neither the transmitter nor the receiver, namely,
\begin{equation}
P_{Y|U}(y|u)=\int_{-\infty}^{+\infty}\mbox{d}a\cdot P_A(a)P_{Y|U,A}(y|u,a).
\end{equation}
where $P_A(a)$ is the density of $A$.
If we think of $I_G(U;Y)$
as a functional of $P_{Y|U}$, denoted $\calI(P_{Y|U}(\cdot|u))$,
then it is a convex functional, namely,
\begin{equation}
\calI(P_{Y|U}(\cdot|u))=
\calI\left(\int_{-\infty}^{+\infty}\mbox{d}a P_A(a)P_{Y|U,A}(\cdot|u,a)\right)\le 
\int_{-\infty}^{+\infty}\mbox{d}a P_A(a)\calI(P_{Y|U,A}(\cdot|u,a)).
\end{equation}
This is a desirable property because the r.h.s.\ reflects a situation where
$A$ is known to both parties, whereas the l.h.s.\ pertains to the
situation where $A$ is unknown, so the lower bound associated with the case
where $A$ is unknown is always tighter than the expectation of the
lower bound pertaining to a known $A$. The WWB, on the other hand, does not
have this convexity property, as we shall see.

Consider now the case where $A$ is a fading parameter, drawn only once and
kept fixed throughout the entire observation time $T$.
More precisely, our model is the same as before except that
now the signal is subjected to fading according to
\begin{equation}
y(t)=a\cdot x(t,u)+n(t),~~~0\le t < T,
\end{equation}
where $a$ and $u$ are realizations of the random variables $A$ and $U$,
respectively. For the sake of convenience in the analysis, we assume that
$A$ is a zero--mean Gaussian random variable with variance $\sigma^2$ (other
densities are, of course, possible too).

We next compare the three corresponding bounds in this case.
The overall channel from  $U$ to $Y$ is
\begin{equation}
P_{Y|U}(y|u)\propto
\int_{-\infty}^{+\infty}\mbox{d}a\cdot\frac{e^{a^2/(2\sigma^2)}}{\sqrt{2\pi\sigma^2}}\cdot
\exp\left\{-\frac{1}{N_0}\int_0^T[y(t)-a\cdot x(t,u)]^2\mbox{d}t\right\}.
\end{equation}
Carrying out the integration, we readily obtain
\begin{equation}
P_{Y|U}(y|u)\propto\exp\left\{\theta
\left[\int_0^Ty(t)x(t,u)\mbox{d}t\right]^2\right\},
\end{equation}
where
\begin{equation}
\theta\eqd\frac{2\sigma^2}{N_0^2(1+2\sigma^2E/N_0)}.
\end{equation}
Thus,
\begin{eqnarray}
-I_G(U;Y)&=&
\bE\left\{\exp\left\{\frac{\theta}{k+1}
\sum_{i=1}^k\left[\int_0^Ty(t)x(t,u_i)\mbox{d}t\right]^2-\right.\right.\nonumber\\
& &\left.\left.\frac{\theta k}{k+1}\left[\int_0^Ty(t)x(t,u_0)\mbox{d}t\right]^2\right\}\bigg|
U=u_0\right\}.
\end{eqnarray}
Upon substituting $y(t)=Ax(t,u_0)+n(t)$, one obtains, after some
straightforward algebra
\begin{eqnarray}
-I_G(U;Y)&=&
\bE\exp\left\{\frac{\theta}{k+1}\left(A^2E^2\sum_{i=1}^k\rho^2(U_0,U_i)+2AE
\sum_{i=1}^k\rho(U_0,U_i)Z_i+\sum_{i=1}^kZ_i^2
\right)-\right.\nonumber\\
& &\left.\frac{\theta k}{k+1}(A^2E^2+2AEZ_0+Z_0^2)\right\},
\end{eqnarray}
where
\begin{equation}
Z_i=\int_0^T n(t)x(t,u_i)\mbox{d}t,~~~~~i=0,1,2,\ldots,k,
\end{equation}
and where the expectation is w.r.t.\ the randomness of $A$, $\{U_i\}$ and
$\{Z_i\}$.
Obviously, given $A$ and $\{U_i\}$, the random variables $\{Z_i\}$ 
are jointly Gaussian with zero--mean with covariances
$\frac{N_0}{2}E\rho(U_i,U_j)$.
Motivated by the discussion at the end of Subsection 4.2,
we now adopt the assumption of signals with rapidly vanishing correlation.
In other words, we assume that $\rho(u,u+h)$ vanishes so rapidly\footnote{Consider an
asymptotic regime under which, the signal $x(t,u)$ depends on an additional
(design) parameter $\Delta$, so that
for every $h\ne 0$, $\rho(h)\to 0$ as $\Delta$ tends to a certain limit, 
and that this
limit is taken {\it before} the limit $E/N_0\to\infty$. 
For example, if
$x(t,u)=x_0(t-u)$ is a rectangular pulse of amplitude $\sqrt{E/\Delta}$ and
duration $\Delta$, then $\rho(h)=[1-|h|/\Delta]_+$ which obviously
vanishes as $\Delta\to 0$ for every $h\ne 0$.}
as a function of $h$ for every $u$, that it is safe to
neglect $\rho(U_i,U_j)$ altogether for all $i\ne j$. This would make $\{Z_i\}$ independent and
simplify the above expression to
\begin{equation}
-I_G(U;Y)=\bE\left[\exp\left\{-\frac{\theta kE^2A^2}{k+1}\right\}
\exp\left\{-\frac{\theta
k}{k+1}(Z_0^2+2AEZ_0)\right\}\right]\cdot
\left(\bE\exp\left\{-\frac{\theta
Z_1^2}{k+1}\right\}\right)^k
\end{equation}
Upon calculating the expectation (w.r.t.\ both $A$ and $\{Z_i\}$), we obtain
\begin{eqnarray}
-I_G(U;Y)&=&\left[\frac{(k+1)(1+2\sigma^2E/N_0)}{k+1+2k\sigma^2E/N_0}\right]^{k/2}\times\nonumber\\
& &\sqrt{\frac{(k+1)(1+2\sigma^2E/N_0)}{(k+1)(1+2\sigma^2E/N_0)+2k\sigma^2E/N_0}}\cdot
\frac{1}{\sqrt{1+2\mu\sigma^2}},
\end{eqnarray}
where
\begin{equation}
\mu\eqd \frac{2k\sigma^2(E/N_0)^2}{2(2k+1)\sigma^2E/N_0+k+1}.
\end{equation}
Considering the high--SNR regime 
($E/N_0 \gg 1$), this is approximated as
\begin{equation}
-I_G(U;Y)\approx\frac{1}{\sqrt{2}}\left(1+
\frac{1}{k}\right)^{(k+1)/2}\cdot\frac{1}{\sigma\sqrt{E/N_0}}\eqd
\frac{f_k}{\sigma\sqrt{E/N_0}}.
\end{equation}
Applying the high--resolution approximation of $D(R)$ for $k\ge 3$, we get:
\begin{equation}
\bE(U-V)^2\ge \frac{g_k}{\sigma}\cdot\sqrt{\frac{N_0}{E}},
\end{equation}
where
\begin{equation}
g_k =\frac{1}{4\sqrt{2}}\left(1-\frac{2}{k}\right)^k
\left(1+
\frac{1}{k}\right)^{(k+1)/2}.
\end{equation}
A simple numerical study indicates that $\{g_k\}$ 
is monotonically increasing and so the best bound is
obtained for $k\to\infty$ (infinitely many replicas), where the constant is:
\begin{equation}
g_\infty=\lim_{k\to\infty}g_k=\frac{1}{4\sqrt{2}}\cdot
e^{-2}\cdot\sqrt{e}=\frac{1}{4\sqrt{2}e^{3/2}}=0.03944.
\end{equation}
Thus, our asymptotic lower bound for high SNR is
\begin{equation}
\liminf_{E/N_0\to\infty}\sqrt{\frac{E}{N_0}}\cdot\bE(U-V)^2\ge
\frac{0.03944}{\sigma}.
\end{equation}
The WWB \cite[p.\ 51]{Weiss85}, in its more general form,
is given by
\begin{equation}
\mbox{WWB}=\sup_{h\ne 0,~s\in[0,1]}\frac{h^2
e^{2\mu(s,h)}}{e^{\mu(2s,h)}+e^{\mu(2-2s,-h)}-2e^{\mu(s,2h)}},
\end{equation}
where 
\begin{equation}
e^{\mu(s,h)}=
\bE\left[\frac{P_{Y|U}(Y|U+h)}{P_{Y|U}(Y|U)}\right]^s,~~~~~~~s\in[0,1]
\end{equation}
which for the fading channel
under the high SNR regime of rapidly vanishing correlation signals,
can be shown (using
similar calculations as above) to be given by
\begin{equation}
\label{musf}
e^{\mu(s,h)}\approx \left\{\begin{array}{ll}
\sqrt{\frac{1+2\sigma^2E/N_0}{(1+2s\sigma^2E/N_0
)(1+2[1-s]\sigma^2E/N_0)}} & h\ne 0\\
1 & h=0\end{array}\right.
\end{equation}
The problem is that, unless $s=1/2$, either $2s > 1$ or $2-2s >
1$, and so correspondingly, for large 
enough values of $E/N_0$, either $e^{\mu(2s,h)}$ or $e^{\mu(2-2s,-h)}$ at
the denominator diverge, and the WWB becomes useless. Thus, the only feasible
choice of $s$ is $s=1/2$, in which case, the WWB becomes
\begin{equation}
\mbox{WWB}=\sup_{h\ne 0}\frac{h^2
e^{2\mu(1/2,h)}}{2[1-e^{\mu(1/2,2h)}]}.
\end{equation}
But $e^{\mu(1/2,h)}$ is exactly our information measure for $k=1$, and so,
\begin{equation}
\mbox{WWB}=\frac{f_1^2N_0/(\sigma^2E)}
{2[1-f_1/(\sigma\sqrt{E/N_0})]}.
\end{equation}
As can be seen, the WWB decays according to $(E/N_0)^{-1}$ rather than
$(E/N_0)^{-1/2}$ and hence inferior to the data processing bound.

The channel--coding bound is based on a universal lower bound on the
probability of error, which holds for every signal set. The problem is
that under fading, we are not aware of such a universal lower bound.
The only remaining alternative then is to use a lower bound corresponding to the case where $A$
is known to the receiver, and then to take the expectation w.r.t.\
$A$, although one might argue that this comparison is not quite fair.
Nonetheless, the derivation of this appears in Appendix E and the result is
\begin{equation}
\liminf_{E/N_0\to\infty}\sqrt{\frac{E}{N_0}}\cdot\bE(U-V)^2\ge
\frac{1}{128\pi\sqrt{2}\sigma}=\frac{0.001758}{\sigma}.
\end{equation}
Thus, the data processing bound is better by a factor of 22.4 (13.5dB).

Yet another comparison, perhaps more fair, can be made with
a related bound, which based on binary hypothesis testing,
but has the advantage of avoiding the use of the Chebychev inequality,
that was used in the channel--coding bound. This is 
the Chazan--Zakai--Ziv bound (CZZB), an improved version of the Ziv--Zakai
bound \cite{ZZ69}. According to the CZZB, applied to our problem (see Appendix
F for the derivation),
\begin{equation}
\liminf_{E/N_0\to\infty}\sqrt{\frac{E}{N_0}}\cdot\bE(U-V)^2\ge
\frac{0.00716}{\sigma},
\end{equation}
which is again significantly smaller than our bound.
Thus, we observe that while the WWB and the CZZB are excellent
bounds for ordinary channels without fading, when it comes to
channels with fading,
the proposed data--processing bound has
an advantage.

\section{Conclusion}

In this work, we have explored a certain class of information
measures \cite{Gurantz74}, which although being a special case of the Zakai--Ziv information
measures \cite{ZZ75}, it has an interesting structure that calls for attention.
We first put this class of information measures in the broader perspective,
relating it to other information measures, like those of \cite{ZZ75}, and
then, by a specific choice of the convex functions, 
we defined a generalized notion of the Chernoff divergence that is 
based on an arbitrary number of replicas of the channel.
Relations have be drawn between the generalized Chernoff divergence and
the Gallager function, the ordinary Chernoff divergence, and even more specifically, the
Bhattacharyya distance. We have also suggested a somewhat more general
structured class based on factor trees. We then applied the data
processing inequality, based on the generalized Chernoff divergence, and
demonstrated that sometimes bounds can be improved by using more than $k+1=2$
replicas. In particular, for the AWGN three replicas is the optimum number in
the AWGN model, thus improving on \cite{ZZ75}, where only two replicas were used
(the ordinary Bhattacharyya distance). While this bound still falls short
compared to other bounds available from estimation theory, the data processing
bound seems to be more powerful than others when it comes to channels
with uncertainty, like fading channels. In this case, the limit of $k\to
\infty$ gives the best result.

\section*{Acknowledgment}

Interesting discussions with Shlomo Shamai are acknowledged with thanks.

\section*{Appendix A}
\renewcommand{\theequation}{A.\arabic{equation}}
    \setcounter{equation}{0}
{\bf Low Resolution Analysis}

Low resolution analysis corresponds to very small values of $s$, which can be
handled by
a first order Taylor series expansion
of the functions $F(s)$, $C(s)$ and $G(s)$. Specifically,
\begin{eqnarray}
C(s)\approx 1+\frac{k+1}{12k}\cdot s\\
F(s)\approx \frac{1}{12}-\frac{k+1}{80k}\cdot s\\
G^{k+1}(s)\approx 1- \frac{k+1}{12k}\cdot s.
\end{eqnarray}
Thus,
\begin{equation}
D_s=C(s)F(s)\approx \frac{1}{12}-\frac{k+1}{180k}\cdot s
\end{equation}
and
\begin{equation}
-R(D_s)=C(s)[G(s)]^{k+1}\approx 1-\left(\frac{k+1}{12k}\right)^2\cdot s^2
\end{equation}
or
\begin{equation}
s\approx \frac{12k}{k+1}\sqrt{R+1}.
\end{equation}
and so
\begin{equation}
D(R)\approx\frac{1}{12}-\frac{1}{15}\sqrt{R+1}.
\end{equation}

\section*{Appendix B}
\renewcommand{\theequation}{B.\arabic{equation}}
    \setcounter{equation}{0}

{\bf High Resolution Analysis}

High resolution corresponds to $s\gg 1$.
In this case, we have
\begin{eqnarray}
\frac{1}{C(s)}&=&\int_{-1/2}^{+1/2}\frac{\mbox{d}w}{(1+sw^2)^{1+1/k}}\nonumber\\
&=&\frac{1}{\sqrt{s}}\int_{-\sqrt{s}/2}^{+\sqrt{s}/2}\frac{\mbox{d}(\sqrt{s}w)}
{(1+(\sqrt{s}w)^2)^{1+1/k}}\nonumber\\
&\approx&\frac{1}{\sqrt{s}}\int_{-\infty}^{+\infty}\frac{\mbox{d}t}
{(1+t^2)^{1+1/k}}\nonumber\\
&\eqd& \frac{c_k}{\sqrt{s}},
\end{eqnarray}
Now, according to the relations between the functions $C$, $F$ and $G$,
derived in Subsection 4.1, we have:
\begin{eqnarray}
G(s)&=&\frac{1}{(1+s/4)^{1/k}}+\frac{2s}{k}F(s)\nonumber\\
&\approx&\frac{4^{1/k}}{s^{1/k}}+\frac{2s}{k}F(s)
\end{eqnarray}
and also
\begin{equation}
G(s)=\frac{1}{C(s)}+sF(s)\approx \frac{c_k}{\sqrt{s}}+sF(s).
\end{equation}
Comparing the two expressions of $G(s)$, we get
\begin{equation}
\frac{c_k}{\sqrt{s}}+sF(s)\approx
\frac{4^{1/k}}{s^{1/k}}+\frac{2s}{k}F(s),
\end{equation}
which leads to the equation
\begin{equation}
\label{feq}
sF(s)\left(1-\frac{2}{k}\right)\approx
\frac{4^{1/k}}{s^{1/k}}-\frac{c_k}{\sqrt{s}}.
\end{equation}
At this stage, we have to handle separately the cases $k=1$, $k=2$ and $k>2$.

Let us consider the case $k=1$ first. In this case, the last equation reads
\begin{equation}
-sF(s)\approx
\frac{4}{s}-\frac{c_1}{\sqrt{s}}\approx -\frac{c_1}{\sqrt{s}}
\end{equation}
and so,
\begin{equation}
F(s)\approx \frac{c_1}{s^{3/2}}.
\end{equation}
Thus, from the distortion equation,
\begin{equation}
D_s=C(s)F(s)\approx \frac{\sqrt{s}}{c_1}\cdot
\frac{c_1}{s^{3/2}}=\frac{1}{s},
\end{equation}
or equivalently, $s=1/D_s$.
Now,
\begin{equation}
G(s)=\frac{1}{C(s)}+sF(s)\approx\frac{c_1}{\sqrt{s}}+s\cdot\frac{c_1}{s^{3/2}}=\frac{2c_1}{\sqrt{s}}.
\end{equation}
From the rate equation, we have
\begin{eqnarray}
-R(D_s)&=& C(s)[G(s)]^2\nonumber\\
&\approx& \frac{\sqrt{s}}{c_1}\cdot \frac{4c_1^2}{s}\nonumber\\
&=& \frac{4c_1}{\sqrt{s}}=4c_1\sqrt{D_s},
\end{eqnarray}
which means
\begin{equation}
R(D)\approx -4c_1\sqrt{D}.
\end{equation}
or equivalently, the distortion--rate function is
\begin{equation}
D(R)\approx\frac{R^2}{16c_1^2},
\end{equation}
where it should be kept in mind that $R$ takes on values in the range $[-1,0]$
in this case.

The case $k=2$ is handled as follows:
\begin{equation}
G(s)=\int_{-1/2}^{+1/2}\frac{\mbox{d}w}{\sqrt{1+sw^2}}=\frac{1}{\sqrt{s}}\ln\frac{\sqrt{s/4+1}+\sqrt{s}/2}
{\sqrt{s/4+1}-\sqrt{s}/2}=\frac{1}{\sqrt{s}}\ln\left(1+\frac{s}{2}+\sqrt{s\left(\frac{s}{4}+1\right)}\right),
\end{equation}
and so, for $s$ large $G(s)\approx (\ln s)/\sqrt{s}$.
By comparing the two expressions for $G(s)$, we find that
$C(s)=\sqrt{1+s/4}\approx \sqrt{s}/2$. Consequently,
\begin{equation}
F(s)=\frac{G(s)-1/C(s)}{s}\approx \frac{\ln s}{s^{3/2}}.
\end{equation}
Thus, $D_s=F(s)C(s)\approx (\ln s)/(2s)$ and $-R(D_s)=C(s)G^3(s)\approx
(\ln^3s)/(2s).$ In the high--resolution limit, the logarithmic terms are
relatively negligible and so, we can deduce that
\begin{equation}
\lim_{s\to\infty}\frac{\log D_s}{\log [-R(D_s)]}=
\lim_{D\to 0}\frac{\log D}{\log [-R(D)]}=1.
\end{equation}
Finally, we examine the case $k > 2$.
Returning to eq.\ (\ref{feq}), now we have:
\begin{equation}
sF(s)\left(1-\frac{2}{k}\right)\approx
\frac{4^{1/k}}{s^{1/k}},
\end{equation}
and so
\begin{equation}
F(s)\approx \frac{k4^{1/k}}{(k-2)s^{1+1/k}}.
\end{equation}
and
\begin{equation}
G(s)\approx\frac{c_k}{\sqrt{s}}+\frac{k4^{1/k}}{(k-2)s^{1/k}}\approx
\frac{k4^{1/k}}{(k-2)s^{1/k}}.
\end{equation}
The distortion equation then gives
\begin{eqnarray}
D_s&=&C(s)F(s)=\frac{\sqrt{s}}{c_k}\cdot
\frac{k4^{1/k}}{(k-2)s^{1+1/k}}\nonumber\\
&=& \frac{k4^{1/k}}{(k-2)c_ks^{1/2+1/k}}\nonumber\\
\end{eqnarray}
and the rate equation yields
\begin{eqnarray}
-R(D_s)&=& C(s)[G(s)]^{k+1}\nonumber\\
&\approx&\frac{\sqrt{s}}{c_k}\cdot\left[
\frac{k4^{1/k}}{(k-2)s^{1/k}}\right]^{k+1}\nonumber\\
&=&\frac{4^{1+1/k}}{c_ks^{1/2+1/k}}\cdot\left(\frac{k}{k-2}\right)^{k+1}\nonumber\\
&=&4\left(\frac{k}{k-2}\right)^k\cdot D_s,
\end{eqnarray}
Thus, the rate--distortion function
and the distortion--rate function are approximated as
\begin{equation}
R(D)\approx-4\left(\frac{k}{k-2}\right)^k\cdot D;
~~~~~~D(R)\approx-\frac{1}{4}\left(1-\frac{2}{k}\right)^k\cdot R,
\end{equation}

\section*{Appendix C}
\renewcommand{\theequation}{C.\arabic{equation}}
    \setcounter{equation}{0}

{\bf Higher Order Moments}

The high--resolution analysis can easily be extended to handle general moments
of the
estimation error, $\bE|U-V|^p$, $p > 0$ ($p$ should not necessarily be
integer). This gives for large $s$,
\begin{equation}
C(s)\approx
\frac{s^{1/p}}{c};~~~c\eqd\int_{-1/2}^{+1/2}\frac{\mbox{d}t}{[1+|t|^p]^{1+1/k}}
\end{equation}
and
\begin{equation}
\left(1-\frac{p}{k}\right)sF(s) \approx
\frac{2^{p/k}}{s^{1/k}}-\frac{c}{s^{1/p}}.
\end{equation}
Here, we have to handle separately the cases $k < p$ and $k > p$ (and the
case $k=p$ will not be covered here, but since $p$ is allowed to be
non--integer, it can be approached by either $p\downarrow k$ or $p\uparrow
k$). In the case $k < p$, we have
\begin{equation}
\left(\frac{p}{k}-1\right)sF(s)\approx \frac{c}{s^{1/p}}
\end{equation}
and so
\begin{equation}
F(s)\approx \frac{kc}{p-k}\cdot\frac{1}{s^{1+1/p}}.
\end{equation}
Thus,
\begin{equation}
D_s=C(s)F(s)=\frac{k}{(p-k)s}.
\end{equation}
Now,
\begin{equation}
G(s)=\frac{1}{C(s)}+sF(s)\approx
\frac{c}{s^{1/p}}+\frac{kc}{(p-k)s^{1/p}}=\frac{pc}{(p-k)s^{1/p}}
\end{equation}
and so
\begin{equation}
-R(D_s)=C(s)[G(s)]^{k+1}\approx
c^k\left(\frac{p}{p-k}\right)^{k+1}\cdot\frac{1}{s^{k/p}}.
\end{equation}
Thus,
\begin{equation}
D(R)\approx S_1(k,p)[-R]^{p/k}.
\end{equation}
where
\begin{equation}
S_1(k,p)=\frac{k}{c^p(p-k)}\cdot\left(1-\frac{k}{p}\right)^{p(1+1/k)}.
\end{equation}
Note that in terms of the asymptotic behavior for small values of $-R$, the
best choice of $k$ is the largest integer strictly less than $p$.
For $p$ integer, this means $k=p-1$.
As for the case $k > p$, we get:
\begin{equation}
\left(1-\frac{p}{k}\right)sF(s)\approx \frac{2^{p/k}}{s^{1/k}}
\end{equation}
or
\begin{equation}
F(s)\approx \frac{k2^{p/k}}{(k-p)s^{1+1/k}}.
\end{equation}
So
\begin{equation}
D_s=C(s)F(s)\approx \frac{k2^{p/k}}{(k-p)cs^{1+1/k-1/p}}
\end{equation}
Here,
\begin{equation}
G(s)=\frac{c}{s^{1/p}}+\frac{k2^{p/k}}{(k-p)s^{1/k}}\approx
c{k2^{p/k}}{(k-p)s^{1/k}}.
\end{equation}
Then,
\begin{equation}
-R(D_s)=C(s)[G(s)]^{k+1}\approx 2^p\left(\frac{k}{k-p}\right)^kD_s,
\end{equation}
and we get
\begin{equation}
D(R)\approx - S_2(k,p)R,
\end{equation}
where
\begin{equation}
S_2(k,p)= 2^{-p}\left(1-\frac{p}{k}\right)^k.
\end{equation}

\section*{Appendix D}
\renewcommand{\theequation}{D.\arabic{equation}}
    \setcounter{equation}{0}

{\bf Derivation of the Channel--Coding Bound}

For a given positive integer $M$, consider the
following chain of inequalities:
\begin{eqnarray}
& &\bE(U-V)^2\nonumber\\
&\ge&
\left(\frac{1}{2M}\right)^2\mbox{Pr}\left\{|U-V|\ge
\frac{1}{2M}\right\}\nonumber\\
&=&\frac{1}{4M^2}\cdot\int_{-1/2}^{+1/2}\mbox{d}u\cdot\mbox{Pr}\left\{|U-V|\ge
\frac{1}{2M}\bigg| U=u\right\}\nonumber\\
&=&\frac{1}{4M^2}\cdot\sum_{i=0}^{M-1}\int_{-1/(2M)}^{+1/(2M)}\mbox{d}u\cdot
\mbox{Pr}\left\{|U-V|\ge
\frac{1}{2M}\bigg| U= \frac{2i+1}{2M}-\frac{1}{2}+ u \right\}\nonumber\\
&=&\frac{1}{4M}\cdot \int_{-1/(2M)}^{+1/(2M)}\mbox{d}u
\cdot\frac{1}{M}\sum_{i=0}^{M-1}\mbox{Pr}\left\{|U-V|\ge
\frac{1}{2M}\bigg| U=\frac{2i+1}{2M}-\frac{1}{2}+ u\right\}.
\end{eqnarray}
Now, note that the integrand of the last expression has a simple
interpretation: Consider the codebook of signals
$\{x(t,u_i)\}$,
$0\le t < T$, $i=0,1,\ldots,M-1$ where
$u_i=(2i+1)/(2M)-1/2+ u$,
and consider the (suboptimum) decoder that first estimates $U$ by an arbitrary
estimator $V$ and then decodes the message according to the $u_i$ that
is nearest to $V$. The integrand in the last line above is simply the
probability of error of that decoder. This probability of error is lower
bounded \cite[p.\ 174, eqs.\ (3.73) and (3.75)]{VO79} according to
\begin{eqnarray}
&&\frac{1}{M}\sum_{i=0}^{M-1}\mbox{Pr}\left\{|U-V|\ge
\frac{1}{2M}\bigg| U=\frac{2i+1}{2M}-\frac{1}{2}+ u
\right\}\nonumber\\
&\ge&
\frac{1}{2}Q\left(\sqrt{\frac{E}{N_0}\cdot\frac{M/2}{M/2-1}}\right)\nonumber\\
&=&\frac{1}{2}Q\left(\sqrt{\frac{E}{N_0}\cdot\frac{M}{M-2}}\right),
\end{eqnarray}
where now $M/2$ should be an integer at least as large as $2$, namely,
$M=4,6,8,\ldots$, Thus,
\begin{equation}
\bE(U-V)^2\ge\frac{1}{4M}\cdot\int_{-1/(2M}^{+1/(2M)}\mbox{d}u
\frac{1}{2}\cdot Q\left(\sqrt{\frac{E}{N_0}\cdot\frac{M}{M-2}}\right)
=\frac{1}{8M^2}\cdot Q\left(\sqrt{\frac{E}{N_0}\cdot\frac{M}{M-2}}\right).
\end{equation}

\section*{Appendix E}
\renewcommand{\theequation}{E.\arabic{equation}}
    \setcounter{equation}{0}

{\bf Channel--Coding Bound for the AWGN Fading Channel}

For a given value of the fading parameter $A=a$, the earlier derivation
of the channel--coding bound implies
\begin{equation}
\bE(U-V)^2\ge \frac{1}{8M^2}\cdot
Q\left(\sqrt{\frac{a^2E}{N_0}\cdot\frac{M}{M-2}}\right).
\end{equation}
Averaging over $A$ and using Craig's formula (see, e.g., \cite{TA99}), we have
\begin{eqnarray}
\bE(U-V)^2&\ge&\frac{1}{8M^2}\int_{-\infty}^{+\infty}\mbox{d}a\frac{e^{-a^2/(2\sigma^2)}}{\sqrt{
2\pi\sigma^2}}
Q\left(\sqrt{a^2\frac{E}{N_0}\cdot\frac{M}{M-2}}\right)\nonumber\\
&=&\frac{1}{8\pi
M^2}\int_{-\infty}^{+\infty}\mbox{d}a\frac{e^{-a^2/(2\sigma^2)}}{\sqrt{
2\pi\sigma^2}}
\int_0^{\pi}\mbox{d}\theta\cdot\exp\left\{-\frac{a^2E
M}{2(M-2)N_0\sin^2\theta}\right\}\nonumber\\
&=&\frac{1}{8\pi M^2}\int_0^{\pi}\mbox{d}\theta
\int_{-\infty}^{+\infty}\mbox{d}a\frac{e^{-a^2/(2\sigma^2)}}{\sqrt{
2\pi\sigma^2}}
\cdot\exp\left\{-\frac{a^2E
M}{2(M-2)N_0\sin^2\theta}\right\}\nonumber\\
&=&\frac{1}{8\pi
M^2}\int_0^{\pi}\frac{\mbox{d}\theta}{\sqrt{1+\sigma^2EM/[(M-2)N_0\sin^2\theta]}}\nonumber\\
&=&\frac{1}{8\pi
M^2}\int_0^{\pi}\frac{\mbox{d}\theta\sin\theta}{\sqrt{\sin^2\theta+E\sigma^2M/[N_0(M-2)]}}\nonumber\\
&>&\frac{1}{8\pi
M^2}\int_0^{\pi}\frac{\mbox{d}\theta\sin\theta}{\sqrt{1+\sigma^2EM/[N_0(M-2)]}}\nonumber\\
&=&\frac{1}{8\pi
M^2\sqrt{1+\sigma^2EM/[N_0(M-2)]}}.
\end{eqnarray}
For $E/N_0$ large, this is approximately,
$$\frac{1}{8\pi
\sigma\sqrt{E/N_0}}\sqrt{\frac{M-2}{M^5}},$$
which is maximized (for even $M > 2$) by $M=4$ to yield
\begin{equation}
\liminf_{E/N_0\to\infty}\sqrt{\frac{E}{N_0}}\cdot\bE(U-V)^2\ge
\frac{1}{128\pi\sqrt{2}\sigma}=\frac{0.001758}{\sigma}.
\end{equation}

\section*{Appendix F}
\renewcommand{\theequation}{F.\arabic{equation}}
    \setcounter{equation}{0}

{\bf Derivation of the Chazan--Zakai--Ziv Bound}

The CZZB \cite{CZZ75} asserts that
\begin{equation}
\bE(U-V)^2\ge \int_0^1
\mbox{d}h\cdot h\int_{-1/2}^{1/2-h}\mbox{d}u\cdot P_e(u,u+h),
\end{equation}
where $P_e(u,u+h)$ is the probability of error associated with optimum
hypothesis testing between the hypotheses $y(t)=Ax(t,u)+n(t)$ and
$y(t)=Ax(t,u+h)+n(t)$, assuming equal priors. Let us denote the probabilities
of error of the two kinds by $P_e(u\to u+h)$ and $P_e(u+h\to h)$. Then,
according to the
Shannon--Gallager--Berlekamp theorem \cite[p.\ 159, Theorem 3.5.1]{VO79},
for every $s\in[0,1]$, at least one of the two following inequalities must
hold:
\begin{eqnarray}
P_e(u\to u+h)&>&\frac{1}{4}\exp[\mu(s,h)-s\mu'(s,h)-s\sqrt{2\mu''(s,h)}]\eqd
A(s)\\
P_e(u+h\to
u)&>&\frac{1}{4}\exp[\mu(s,h)+(1-s)\mu'(s,h)-(1-s)\sqrt{2\mu''(s,h)}]\eqd B(s)
\end{eqnarray}
where $\mu'(s,h)$ and $\mu''(s,h)$ denote the first two partial derivatives of
$\mu(s,h)$ w.r.t\ $s$, and where for rapidly--vanishing--correlation 
signals, $\mu(s,h)$ is given by the (first line of) eq.\
(\ref{musf}). Since $\mu(1/2,h)=\ln[f_1/(\sigma\sqrt{E/N_0})$, 
$\mu'(1/2,h)=0$ and $\mu''(1/2)\approx 1/4$ at the high SNR
limit, this implies that 
\begin{eqnarray}
P_e(u,u+h)&=&\frac{P_e(u\to u+h)+P_e(u+h\to u)}{2}\nonumber\\
&>& \sup_{0\le s\le 1}\frac{1}{2}\min\{A(s),B(s)\}\nonumber\\
&\ge&\frac{1}{2}\min\{A(1/2),B(1/2)\}\nonumber\\
&=&\frac{1}{8}\exp\{\mu(1/2,h)-0.5\cdot\sqrt{2\mu''(1/2,h)}\}\nonumber\\
&\approx&\frac{1}{8e^{\sqrt{2}}}\cdot\frac{f_1}{\sigma\sqrt{E/N_0}}\nonumber\\
&=&\frac{0.042977}{\sigma\sqrt{E/N_0}},
\end{eqnarray}
and so,
\begin{equation}
\bE(U-V)^2\ge
\frac{0.042977}{\sigma\sqrt{E/N_0}}\int_0^1h(1-h)\mbox{d}h
=\frac{0.00716}{\sigma\sqrt{E/N_0}},
\end{equation}

\end{document}

%% file: factortree.pstex_t
\begin{picture}(0,0)%
\includegraphics{factortree.pstex}%
\end{picture}%
\setlength{\unitlength}{3947sp}%
\begingroup\makeatletter\ifx\SetFigFont\undefined%
\gdef\SetFigFont#1#2#3#4#5{%
  \reset@font\fontsize{#1}{#2pt}%
  \fontfamily{#3}\fontseries{#4}\fontshape{#5}%
  \selectfont}%
\fi\endgroup%
\begin{picture}(2792,2264)(423,-1634)
\put(2875,-747){\makebox(0,0)[lb]{\smash{{\SetFigFont{9}{10.8}{\rmdefault}{\mddefault}{\itdefault}{$Q_1$}%
}}}}
\put(2164,-1174){\makebox(0,0)[lb]{\smash{{\SetFigFont{9}{10.8}{\rmdefault}{\mddefault}{\itdefault}{$L_{c,a}$}%
}}}}
\put(2199,-357){\makebox(0,0)[lb]{\smash{{\SetFigFont{9}{10.8}{\rmdefault}{\mddefault}{\itdefault}{$L_{b,a}$}%
}}}}
\put(1524,-1601){\makebox(0,0)[lb]{\smash{{\SetFigFont{9}{10.8}{\rmdefault}{\mddefault}{\itdefault}{$Q_3$}%
}}}}
\put(1453, 34){\makebox(0,0)[lb]{\smash{{\SetFigFont{9}{10.8}{\rmdefault}{\mddefault}{\itdefault}{$Q_2$}%
}}}}
\put(494,532){\makebox(0,0)[lb]{\smash{{\SetFigFont{9}{10.8}{\rmdefault}{\mddefault}{\itdefault}{$L_{d,b}$}%
}}}}
\put(423,-570){\makebox(0,0)[lb]{\smash{{\SetFigFont{9}{10.8}{\rmdefault}{\mddefault}{\itdefault}{$L_{e,b}$}%
}}}}
\put(487,-1581){\makebox(0,0)[lb]{\smash{{\SetFigFont{9}{10.8}{\rmdefault}{\mddefault}{\itdefault}{\color[rgb]{0,0,0}$L_{f,c}$}%
}}}}
\end{picture}%

%% file: p146.bbl
\begin{thebibliography}{AA}

\bibitem{Andelman74}
D.~Andelman, {\it Bounds According to a Generalized Data Processing Theorem},
M.Sc.\ final paper (in Hebrew), Department of Electrical Engineering, Technion
-- Israel Institute of Technology, Haifa, Israel, October 1974.

\bibitem{Arimoto73}
S.~Arimoto, ``On the converse to the coding theorem for discrete 
memoryless channels'',
{\em IEEE Transactions on Information Theory\/},
pp.~357--359, May 1973.

\bibitem{BV04}
S.~Boyd and L.~Vandenberghe, {\it Convex
Optimization}, Cambridge University Press, 2004.

\bibitem{CZZ75}
D.~Chazan, M.~Zakai, and J.~Ziv, ``Improved lower bounds on 
signal parameter estimation,''
{\it IEEE Trans.\ Inform.\ Theory}, vol.\ IT--21, no.\ 1, pp.\ 90--93,
January 1975.

\bibitem{CT06} 
T.~M.~Cover and J.~A.~Thomas,
{\em Elements of Information Theory\/}.
John Wiley \& Sons, Second Edition, Hoboken NJ, USA, 2006.

\bibitem{Csiszar63}
I.~Csisz\'ar, ``Eine informationstheoretische Ungleichung und ihre Anwendung
auf den Beweis der Ergodizit\"at von Markoffschen Ketten,'' {\it Publ.\ Math.\
Inst.\ Hungar.\ Acad.}, vol.\ 8, pp.\ 95--108, 1963.

\bibitem{Csiszar72}
I.~Csisz\'ar, ``A class of measures of informativity of observation
channels,'' {\it Periodica Mathematica Hungarica}, vol.\ 2 (1--4),
pp.\ 191--213, 1972.

\bibitem{CS04}
I.~Csisz\'ar and P.~Shields, ``Information theory and statistics: a
tutorial,'' {\it Foundations and Trends in Communications and Information
Theory}, vol.\ 1, no.\
4, 417--528, 2004.

\bibitem{Gallager68}
R.~G.~Gallager, {\it Information Theory and Reliable Communication},
J.~Wiley \& Sons, 1968.

\bibitem{Gurantz74}
I.~Gurantz, {\it Application of a Generalized Data Processing Theorem},
M.Sc.\ final paper (in Hebrew), Department of Electrical Engineering, Technion -- Israel
Institute of Technology, Haifa, Israel, August 1974.

\bibitem{Hen02}
I.~Hen, {\it The Threshold Effect in the Estimation
of Chaotic Sequences},
M.Sc. dissertation, Department of Electrical Engineering, Technion -- Israel
Institute of Technology, Haifa, Israel,
February 2002.

\bibitem{HM04}
I.~Hen and N.~Merhav, ``On the threshold effect in
the estimation of chaotic sequences,''
{\it IEEE Trans.\ Inform.\ Theory}, vol.\ 50,
no.\ 11, pp.\ 2894--2904, November 2004.

\bibitem{KS93}
G.~Kaplan and S.~Shamai (Shitz), ``Information rates and error exponents
of compound channels with application to antipodal signaling in a fading
environment,'' {\it AE\"U}, vol.~ 47, no.\ 4, pp.\ 228--239, 1993.

\bibitem{Merhav11a}
N.~Merhav, ``Data processing theorems and the second law of
thermodynamics,''
{\it IEEE Trans.\ Inform.\ Theory}, vol.\ 57, no.\ 8, pp.\
4926--4939, August 2011.

\bibitem{Merhav11b}
N.~Merhav, ``Threshold effects in parameter estimation
as phase transitions in statistical mechanics,'' to appear in
{\it IEEE Trans.\ Inform.\ Theory}, October 2011.

\bibitem{TA99}
C.~Tellambura and A.~Annamalai, ``Derivation of Craig's formua for Gaussian
probability function,'' {\it Electronic Letters}, 
vol.\ 35, no.\ 17, pp.\ 1424--1425, August 19, 1999.

\bibitem{VO79}
A.~J.~Viterbi and J.~K.~Omura, {\it Principles of Digital Communication and
Coding}, McGraw--Hill, 1979.

\bibitem{Weiss85}
A.~J.~Weiss, {\it Fundamental Bounds in Parameter Estimation},
Ph.D.\ dissertation, Tel Aviv University, Tel Aviv, Israel, June 1985.

\bibitem{WW85a}
A.~J.~Weiss and E.~Weinstein, ``A lower bound on the mean square error
in random parameter estimation,''
{\em IEEE Transactions on Information Theory\/},
vol.~IT--31, no.~5, pp.~680--682, September 1985.

\bibitem{WW85b}
E.~Weinstein and A.~J.~Weiss, ``Lower bounds on the mean square 
estimation error,'' {\it Proc.\ IEEE}, vol.\ 73, no.\ 9, pp.\ 1433--1434,
September 1985.

\bibitem{WJ65}
J.~M.~Wozencraft and I.~M.~Jacobs, {\it Principles of Communication
Engineering}, John Wiley \& Sons, 1965. Reissued by Waveland Press, 1990.

\bibitem{ZZ75}
M.~Zakai and J.~Ziv, ``A generalization of the rate-distortion theory and
applications,'' in: {\em Information Theory New Trends and Open Problems\/},
edited by G. Longo, Springer-Verlag, 1975, pp.~87--123.

\bibitem{ZZ69}
J.~Ziv and M.~Zakai, ``Some lower bounds on signal parameter 
estimation,'' {\em IEEE Transactions on Information Theory\/},
vol.~IT--15, no.~3, pp.~386--391, May 1969.

\bibitem{ZZ73}
J.~Ziv and M.~Zakai, ``On functionals satisfying a data-processing 
theorem,'' {\em IEEE Trans.~Inform.~Theory\/},
vol.~IT--19, no.~3, pp.~275--283, May 1973.

\end{thebibliography}
